\documentclass[12pt]{article}
\usepackage{epsf}
\usepackage{amsfonts}

\newcounter{subequation}[equation]

\newcommand{\al}{\ensuremath{\alpha}}
\newcommand{\ga}{\ensuremath{\gamma}}
\newcommand{\Ga}{\ensuremath{\Gamma}}
\newcommand{\de}{\ensuremath{\delta}}
\newcommand{\De}{\ensuremath{\Delta}}
\newcommand{\ep}{\ensuremath{\epsilon}}
\newcommand{\vep}{\ensuremath{\varepsilon}}

\newcommand{\la}{\ensuremath{\lambda}}
\newcommand{\om}{\ensuremath{\omega}}
\newcommand{\Om}{\ensuremath{\Omega}}
\newcommand{\p}{\ensuremath{\phi}}
\newcommand{\s}{\ensuremath{\sigma}}
\renewcommand{\S}{\ensuremath{\Sigma}}

\newcommand{\D}{\ensuremath{{\cal D}}}

\renewcommand{\L}{\ensuremath{{\cal L}}}
\newcommand{\M}{\ensuremath{{\cal M}}}

\newcommand{\del}{\ensuremath{\partial}}

\newcommand{\td} {\ensuremath{\tilde}}
\newcommand{\inv}{\ensuremath{^{-1}}}
\newcommand{\half}{\ensuremath{\frac{1}{2}}}
\newcommand{\third}{\ensuremath{\frac{1}{3}}}
\newcommand{\quarter}{\ensuremath{\frac{1}{4}}}
\newcommand{\ul}{\ensuremath{\underline}}

\newcommand{\be}{\begin{equation}}
\newcommand{\ee}{\end{equation}}

\newcommand{\Dps}{D$p$-branes}
\newcommand{\Ds}{D-branes}

\makeatletter

\def\thesubequation{\theequation\@alph\c@subequation}
\def\@subeqnnum{{\rm (\thesubequation)}}
\def\slabel#1{\@bsphack\if@filesw {\let\thepage\relax
   \xdef\@gtempa{\write\@auxout{\string
      \newlabel{#1}{{\thesubequation}{\thepage}}}}}\@gtempa
   \if@nobreak \ifvmode\nobreak\fi\fi\fi\@esphack}
\def\subeqnarray{\stepcounter{equation}
\let\@currentlabel=\theequation\global\c@subequation\@ne
\global\@eqnswtrue
\global\@eqcnt\z@\tabskip\@centering\let\\=\@subeqncr
$$\halign to \displaywidth\bgroup\@eqnsel\hskip\@centering
  $\displaystyle\tabskip\z@{##}$&\global\@eqcnt\@ne
  \hskip 2\arraycolsep \hfil${##}$\hfil
  &\global\@eqcnt\tw@ \hskip 2\arraycolsep
  $\displaystyle\tabskip\z@{##}$\hfil
   \tabskip\@centering&\llap{##}\tabskip\z@\cr}
\def\endsubeqnarray{\@@subeqncr\egroup
                     $$\global\@ignoretrue}
\def\@subeqncr{{\ifnum0=`}\fi\@ifstar{\global\@eqpen\@M
    \@ysubeqncr}{\global\@eqpen\interdisplaylinepenalty \@ysubeqncr}}
\def\@ysubeqncr{\@ifnextchar [{\@xsubeqncr}{\@xsubeqncr[\z@]}}
\def\@xsubeqncr[#1]{\ifnum0=`{\fi}\@@subeqncr
   \noalign{\penalty\@eqpen\vskip\jot\vskip #1\relax}}
\def\@@subeqncr{\let\@tempa\relax
    \ifcase\@eqcnt \def\@tempa{& & &}\or \def\@tempa{& &}
      \else \def\@tempa{&}\fi
     \@tempa \if@eqnsw\@subeqnnum\refstepcounter{subequation}\fi
     \global\@eqnswtrue\global\@eqcnt\z@\cr}
\let\@ssubeqncr=\@subeqncr
\@namedef{subeqnarray*}{\def\@subeqncr{\nonumber\@ssubeqncr}\subeqnarray}
\@namedef{endsubeqnarray*}{\global\advance\c@equation\m@ne%
                           \nonumber\endsubeqnarray}

\makeatother

\begin{document}

\bibliographystyle{unsrt}

\vspace*{-.6in}
\thispagestyle{empty}
\begin{flushright}
DAMTP-1999-97\\
hep-th/9908018
\end{flushright}
\baselineskip = 20pt

\vspace{.5in}
{\Large\bfseries
\begin{center}
Ricci--Flat Branes
\end{center}}
\vspace{.4in}

\begin{center}
D. Brecher\footnote{email d.r.brecher@damtp.cam.ac.uk.} and
M. J. Perry\footnote{email malcolm@damtp.cam.ac.uk.}\\
\emph{D.A.M.T.P., University of Cambridge, Cambridge CB3 9EW, U.K.}
\end{center}
\vspace{1in}

\begin{center}
\textbf{Abstract}
\end{center}
\begin{quotation}
Up to overall harmonic factors, the D8-brane solution of the massive type IIA supergravity
theory is the product of nine--dimensional Minkowski space (the
worldvolume) with the
real line (the transverse space).  We show that the equations of motion allow for
the worldvolume metric to be
generalised to an arbitrary Ricci--flat one.  If this nine--dimensional
Ricci--flat manifold admits Killing spinors, then the resulting
solutions are supersymmetric and satisfy the usual Bogomol'nyi
bound, although they preserve fewer than the usual one half of the
supersymmetries.  We describe the possible choices of such
manifolds, elaborating
on the connection between the existence of Killing spinors and the
self--duality condition on the curvature two--form.  Since the
D8-brane is a domain wall in ten dimensions, we are led to consider the general
case: domain walls in any supergravity theory.  Similar
considerations hold here also.  Moreover, it is shown that the worldvolume of
any magnetic brane --- of which the domain walls are a specific
example --- can be generalised in precisely the same way.  The general class of
supersymmetric solutions have gravitational instantons as their
spatial sections.  Some mention is made of the worldvolume solitons of
such branes.
\end{quotation}
\vfil

\newpage

\pagenumbering{arabic}

\section{Introduction}

The discovery of branes in string theory has brought about some
revolutionary advances in recent years.  From what was initially a
purely perturbative theory, it has been possible
to extract non--perturbative results which have led to new and deep
insights into the nature of string theory.  Perhaps the most
far--reaching of these has been the discovery of string theory's
eleven--dimensional origin, `M-Theory', with its low--energy
limit of eleven--dimensional supergravity.

Ignoring for the time being the D8-brane, the ten--dimensional type IIA and IIB supergravity
theories have solutions which describe the various $p$-branes of type
IIA and IIB string theory.  The field content of the supergravity theories is
precisely what is needed to imply the existence of such branes:
the relevant $(p+1)$--form potentials are all
present~\cite{H2}.  The fundamental string and its magnetic
dual, the 5-brane, couple to the Neveu--Schwarz--Neveu--Schwarz
(NS-NS) potentials; these being present in both the type IIA and IIB
theories.  In addition to these are the \Dps, which couple to the
Ramond--Ramond (R-R) $(p+1)$--form potentials: the type IIA theory has $p$ even,
and type IIB has $p$ odd.  For $p < 3$, the branes couple to an
electric field strength --- these are the fundamental, or electric,
branes.  In $D$ dimensions, Hodge duality allows for a $(p+2)$--form
field strength to be
interchanged with a $(D-(p+2))$--form field strength, so the branes
with $p > 3$ couple to the dual, magnetic field strength.  These are
the solitonic, or magnetic, branes.  (The D3-brane is a special case,
being self--dual in ten dimensions.)  All such branes preserve one--half of the
spacetime supersymmetries.  Moreover, they all have a clear and
well--defined eleven--dimensional origin.

The only aspect of this unified picture which is somewhat unclear is
that of the D8-brane of type IIA string theory.  This is a domain
wall in ten dimensions, which should couple to a ten--form field
strength or, by Hodge duality, a scalar.  As first pointed out by
Polchinski~\cite{P}, the once relatively obscure generalisation of the
type IIA supergravity theory found by Romans~\cite{R} --- the massive IIA theory ---
has the necessary field content~\cite{BRGPT}.  Since type IIA string theory
necessarily includes a D8-brane, it would seem that it is the Romans theory which is
the natural low--energy limit of type IIA string theory.

The Romans theory has some unusual properties: ten--dimensional
Minkowski space is \emph{not} a solution.  Indeed,
none of the Kaluza--Klein compactifications of the theory originally
considered by Romans~\cite{R} are supersymmetric.  This is in contrast
to the D8-brane solution~\cite{PW,BRGPT}
which preserves the usual one half of the
ten--dimensional supersymmetries.  Moreover, the eleven--dimensional origin of
the Romans theory is quite mysterious (although see~\cite{H}).  If the theory is to find a
place within M-Theory, then presumably the D8-brane
would be the dimensional reduction of an M9-brane~\cite{BS}.  It is
unclear, however, how this works in detail.

Although we do not
consider it here, mention should be made of
an alternative ten--dimensional massive supergravity theory, that of Howe, Lambert and
West~\cite{HLW}.  Unlike the Romans theory, this does have a
well--defined eleven--dimensional origin~\cite{HLW,LLP}, but its
connection with string theory is unclear: the NS-NS two--form
which couples to the fundamental strings can be gauged away, so it
would seem that this theory does not contain such strings at all.

All the p-brane solutions of the various supergravity theories,
including the D8-brane, have a common form: in the Einstein frame the
line element is
\be
ds^2 = H^{-\frac{4\td{d}}{\De(D-2)}} d\ul{x} \cdot d\ul{x} +
H^{\frac{4d}{\De(D-2)}} d\ul{y} \cdot d\ul{y},
\label{intro}
\ee

\noindent where $\{x^i\}$ and $\{y^{\al}\}$ are the worldvolume and transverse
coordinates respectively.  The function $H(r)$ is harmonic on the
transverse space, $r$ being the radial coordinate in these
directions.  $d=p+1$, $\td{d} = D-d-2$ denotes the number of
dimensions of the dual brane and, for D-branes and M-branes, we have
$\De=4$.

Now we can ask the question whether it is possible to generalise the
metrics on the worldvolume and transverse space in these
solutions.  That is, whether the field
equations, and the supersymmetry conditions, admit general metrics
instead of the flat ones appearing above.  Indeed, there has been some work to
show that the transverse space of the electric branes can be
generalised, as long as the metric is
Ricci--flat~\cite{DLPS,GGPT,BFK}.  The resulting solution will be
supersymmetric if and only if the transverse space admits Killing
spinors.

In the following section of this paper, we show that the \emph{worldvolume}, as opposed to the
transverse space, of the D8-brane solution can be generalised in just such a
manner.  That is, the solution (\ref{intro}) with $p=8$ and $D=10$ can be generalised so as to
include a non--trivial Ricci--flat worldvolume metric.  As long as
this metric admits Killing spinors, the solution is supersymmetric,
and satisfies the usual Bogomol'nyi bound, as we will show in
subsection 2.2.  We hope that such considerations will shed some light
on the eleven--dimensional origin of the D8-brane.  There are
numerous examples in the literature of possible supersymmetric manifolds
which could be taken as the worldvolume of the D8-brane; and we
discuss some of these in subsections 2.3 and 2.4.  This involves a
consideration of holonomy groups.  The
spatial sections of the D8-brane can take the form of
eight--dimensional gravitational instantons --- manifolds with
self--dual curvature --- of holonomy $\mbox{Spin}(7)$.  Section three is concerned with the
dimensional reduction of the D8-brane, a 7-brane domain wall in nine
dimensions.  Here we show that the manifolds of $G_2$ holonomy described in
the literature are also self--dual; and that these
can be taken as possible spatial sections of this 7-brane.

Since the D8-brane is a domain wall, we are led to consider whether
the same sort of generalisations can be made for domain walls in
\emph{any} supergravity theory.  We show, in section four, that this is
indeed the case.  In particular, it is possible for the 3-brane domain
walls of Ho\v{r}ava--Witten theory to have general Ricci--flat
worldvolumes.  In section five we extend the analysis
to arbitrary magnetic branes, of which the domain wall is but a
special case.  The general statement is, then, that all magnetic
branes can have Ricci--flat worldvolumes.  We concentrate on the
M5-brane here.  Finally, some mention is made of the worldvolume
solitons of such magnetic branes; those of the D4-brane in particular.

The conventions we use are as follows.  The signature of the
metric is $(-, +, \ldots, +)$, the sign of the Riemann curvature tensor is
defined by $R_{abcd} X^d = [ D_a , D_b ] X_c$, and the gamma matrices satisfy $\{ \Ga^a,
\Ga^b \} = 2 g^{ab}$.  As to indices, we take $a,b = 0, \ldots, D-1$
to denote spacetime directions,  $i,j = 0, \ldots, p$ to denote worldvolume directions, and
$\al, \beta = p+1, \ldots, D-1$ to denote transverse directions.   We
are mainly concerned with the case $D=10$ and $p=8$.  Where necessary, we use
underlined indices to denote an (pseudo--)orthonormal
basis and we work exclusively in the Einstein frame.  In the following
section, we use the notation of forms as in~\cite{BRGPT}.  That is, a $q$--form $Q$ has
components $Q_{a_1 \ldots a_q}$ given by
\[
Q = Q_{a_1 \dots a_q} dx^{a_1} \wedge \ldots \wedge dx^{a_q},
\]

\noindent and $|Q|^2 = Q_{a_1 \dots a_q} Q^{a_1 \dots a_q}$.

\section{The Romans Theory}

The ten--dimensional type IIA supergravity theory contains the metric,
the dilaton $\p$
and a two--form $B$ in the NS-NS sector, and a one--form potential $A$
and a three--form potential $C$ in the R-R sector.  The massive type
IIA theory found by Romans~\cite{R} is constructed by allowing the
two-form to `eat' the one--form, thereby generating a
\emph{massive} two--form via a generalised Higgs mechanism.  With the
field strengths given by
\begin{eqnarray}
F &=& 4 dC + 6m(B)^2,\nonumber \\
H &=& 3 dB,\nonumber
\end{eqnarray}

\noindent the Lagrangian for the bosonic sector as given
in~\cite{BRGPT} is
\[
\L = \sqrt{-g} \left( R - \half |\del \p|^2 - \third e^{-\p}
|H|^2 - \frac{1}{12} e^{\half \p} |F|^2 - m^2 e^{\frac{3}{2}\p} |B|^2 -
\half m^2 e^{\frac{5}{2}\p} \right)
\]
\be
+\frac{1}{9} \vep \left( dCdCB + m dC(B)^3 + \frac{9}{20}m^2(B)^5
\right),
\label{lagrangian}
\ee

\noindent where $m$ is the mass of the two--form $B$.  The potential,
or `cosmological term', is of Liouville type, given by
\be
V(\p) = \half m^2 e^{\frac{5}{2}\p}.
\ee

In ten dimensions, we can consider a ten--form
field strength which, by the equations of motion, must have a constant
zero--form Hodge dual.  This constant is just the mass parameter,
$\star F_{[10]} \sim m$, and the Romans theory can
be rewritten to include explicitly the required nine--form
potential~\cite{BRGPT}.  This is an R-R potential, as required to couple to the
D8-brane~\cite{P}, which can be seen by transforming to the string frame, $g_{ab}^{(S)}
= e^{\half \p} g_{ab}^{(E)}$.  The potential is then just a constant
\be
V = \half m^2,
\ee

\noindent the absence of a dilaton factor
indicating the R-R nature of the field.

The D8-brane provides a
natural, and non--trivial, background in Type IIA string theory.
Indeed, the Romans theory has
`massive' fundamental string solutions, coupling to the massive
$B$ field, in just such a background~\cite{JMO}: a solution
which describes the intersection of a fundamental string with a
D8-brane over a D0-brane.  This solution reduces to the usual string
solution of the type IIA theory if the mass parameter $m=0$.

\subsection{Ricci--Flat D8-branes}

Since we are considering D8-branes alone, we turn off all gauge fields.  The
equations of motion following from the Lagrangian (\ref{lagrangian}) are then
\begin{subeqnarray}
R_{ab} &=& \half \del_a \p \del_b \p + \frac{1}{16} m^2
e^{\frac{5}{2}\p} g_{ab},\label{eoma}\\
D^2 \p &=& \frac{5}{4} m^2 e^{\frac{5}{2}\p},\label{eomb}
\end{subeqnarray}

\noindent where $D_a$ is the covariant derivative with respect to the
metric $g_{ab}$.  Note that due to
the Liouville form of the potential, the
dilaton equation (\ref{eoma}b) implies that $\p$ cannot be a constant.  

A manifestly conformally flat D8-brane solution was first discussed
in~\cite{PW}.  A more convenient, but equivalent, one was given in~\cite{BRGPT}; in the Einstein
frame, it is
\be
\left.\begin{array}{ccl} ds^2 &=& H^{\frac{1}{8}} d\ul{x} \cdot d\ul{x} +
H^{\frac{9}{8}} dy^2,\\ [.07in]
e^{\p} &=& H^{-\frac{5}{4}}, \end{array}\right\}
\ee

\noindent where $\{x^i\}$ are the worldvolume coordinates.  $H(y)$ is harmonic on the
single transverse direction $y$.  The
precise form of this function
has implications for the singularity structure of the metric as
discussed in~\cite{LPT}.  With 
\be
H(y) = 1 + m \left| y - y_{0} \right|,
\ee

\noindent the spacetime is free of curvature singularities which would
otherwise be present if we dropped the constant.  There is
still a delta--function singularity at $y = y_0$ however, although
this can be removed, as usual, by adding a source term.  The solution
then describes a D8-brane situated at $y=y_0$.  Note that the mass parameter
$m$ can be positive or negative here.  Performing the coordinate
transformation
\[
d \td{y}^2 = H^{\frac{9}{8}} d y^2,
\] 

\noindent we have
\be
\left.\begin{array}{ccl} ds^2 &=& \left(1 + \frac{25}{16} m |y| \right)^{\frac{2}{25}} d\ul{x} \cdot
d\ul{x} + dy^2,\\ [.07in]
e^{\p} &=& \left( 1 + \frac{25}{16} m |y|
\right)^{-\frac{4}{5}}. \end{array}\right\}
\label{flatcase}
\ee

Consider a generalisation of this solution, taking
\be
ds^2 = F^2 \ga_{ij}(x) dx^i dx^j + dy^2,
\ee

\noindent as our ansatz for the metric.  We also take $F = F(y)$ and $\p = \p(y)$.  The
zehnbeins are $e^{\ul{y}} = dy$ and
$e^{\ul{i}} = F \hat{e}^{\ul{i}}(x)$, where
$\hat{e}^{\ul{i}}$ is the neunbein for the metric $\ga_{ij}$.
The components of the Ricci tensor are
\[
R_{yy} = -9 \frac{F''}{F},
\]
\[
R_{ij} = \hat{R}_{ij} - \ga_{ij} \left( F F'' + 8 F'^2 \right),
\]

\noindent where $\hat{R}_{ij}$ is the Ricci tensor constructed from
the metric $\ga_{ij}$ and a prime denotes a derivative with respect to
$y$.  The non--trivial components of (\ref{eoma}a,b) are then
\begin{subeqnarray}
-18 \frac{F''}{F} &=& \p'^2 +
\frac{1}{8} m^2 e^{\frac{5}{2}\p},\label{curvedeoma}\\
\hat{R}_{ij} &=& \ga_{ij} \left( F F'' + 8F'^2 + \frac{1}{16} F^2 m^2
e^{\frac{5}{2}\p} \right),\label{curvedeomb}\\
\p'' + 9 \frac{F'}{F} \p' &=& \frac{5}{4} m^2
e^{\frac{5}{2}\p}.\label{curvedeomc}
\end{subeqnarray}

\noindent The equations (\ref{curvedeoma}a,c) are solved by
\be
\left.\begin{array}{ccl}
F(y) &=& \left( 1 + \frac{25}{16} m |y - y_0| \right)^{\frac{1}{25}},\\ [.07in]
e^{\p} &=& F(y)^{-20},\end{array}\right\}
\ee

\noindent in which case, the final equation (\ref{curvedeoma}b) becomes
\be
\hat{R}_{ij} = 0.
\ee

\noindent Thus, for any Ricci--flat metric $\ga_{ij}(x)$,
\be
\left.\begin{array}{ccl} ds^2 &=& F^2 \ga_{ij}(x) dx^i dx^j + dy^2,\\ [.07in]
e^{\p} &=& F^{-20},\\ [.07in]
F(y) &=& \left( 1 + \frac{25}{16} m |y - y_0| \right)^{\frac{1}{25}},
\end{array}\right\}
\label{soln}
\ee

\noindent is a solution of the Romans theory.  The interpretation is
obvious: it describes a D8-brane with Ricci--flat
worldvolume.  The flat case (\ref{flatcase}) is then a special case of
this more general family of solutions.  It should be noted, however,
that the isometry group may no longer be the nine--dimensional
Poincar\'{e} group.  Since the overall harmonic function is unchanged, this
generalisation has not altered the
singularity structure of the D8-brane, as long as the metric
$\ga_{ij}$ does not have any singularities itself.  As an example of a
solution which does alter the singularity structure, the worldvolume
of the D8-brane could take the form of a nine--dimensional
Schwarzschild black hole, although this solution will not be
supersymmetric.

\subsection{Supersymmetric D8-Branes and the Bogomol'nyi Bound}

Ricci--flatness of $\ga_{ij}$ is a necessary condition for the
solutions (\ref{soln}) to be supersymmetric.  To see this, consider the supersymmetry
transformations of the spacetime fermionic fields~\cite{R}
\begin{subeqnarray}
\de \psi_a &=& \D_a \ep = \left( D_a - \frac{1}{32} m e^{\frac{5}{4}\p}
\Ga_a \right) \ep,\label{susya}\\
\de \la &=& -\frac{1}{2\sqrt{2}} \left( \Ga^a \del_a \p  + \frac{5}{4} m
e^{\frac{5}{4}\p} \right) \ep,\label{susyb}
\end{subeqnarray}

\noindent where $\ep$ is an arbitrary 32--component Majorana spinor.  Setting
the dilatino variation (\ref{susya}b) to zero gives
\be
\left( \mathbf{1} \mp \Ga_{\ul{y}} \right) \ep = 0,
\label{chiral}
\ee

\noindent so $\ep$ must have a definite chirality, in the sense that $\Ga_{\ul{y}}
\ep = \pm \ep$.  This removes one half of the components of $\ep$.
The sign here is set by the sign of $(y-y_o)$, so the chirality
of $\ep$ changes as we pass through the location of the brane~\cite{BRGPT}.  The
vanishing of the gravitino variation (\ref{susya}a) implies
\begin{subeqnarray}
\ep' &=& \frac{1}{32} m F^{-25} \Ga_{\ul{y}} \ep,\label{susyreqa}\\
\hat{D}_i \ep &=& 0,\label{susyreqb}
\end{subeqnarray}

\noindent where $\hat{D}_i$ is the covariant derivative with respect
to the metric $\ga_{ij}$.  A solution of (\ref{susyreqa}a) is
\be
\ep = F(y)^{1/2} \hat{\ep}(x),
\ee

\noindent where $\hat{\ep}(x)$ is an $SO(8,1)$ `worldvolume spinor'
which does not depend on $y$.

The integrability condition of the remaining equation (\ref{susyreqa}b) is, as usual,
\be
[ \hat{D}_i, \hat{D}_j ] \hat{\ep} = \hat{R}_{ijkl} \hat{\Ga}^{kl} \hat{\ep} = 0,
\label{int}
\ee

\noindent where $\hat{\Ga}^i = F \Ga^i$ satisfy $\{\hat{\Ga}^i , \hat{\Ga}^j\}
= 2 \ga^{ij}$.  By contracting (\ref{int}) with $\hat{\Ga}^j$, we find 
\be
\hat{R}_{ij} = 0,
\label{ricci}
\ee

\noindent a necessary condition for the solutions
(\ref{soln}) to be supersymmetric.  

This can also be seen by a consideration of the integrability
condition of the equation $\de \psi_a =
\D_a \ep = 0$.  This is 
\be
\left[ \D_a, \D_b \right] \ep = \left( R_{abcd} + \frac{1}{128} m^2
e^{\frac{5}{2}\p} g_{ac} g_{bd} \right) \Ga^{cd} \ep + \frac{5}{32} m
e^{\frac{5}{4}\p} \left( \Ga_a \del_b \p - \Ga_b \del_a \p \right) \ep
= 0.
\label{int2}
\ee

\noindent The $\{y,y\}$ component of (\ref{int2}) is trivial.  The
$\{y,i\}$ component is just the chirality condition (\ref{chiral});
and the $\{i,j\}$ component gives
\[
\hat{R}_{ijkl} \hat{\Ga}^{kl} \hat{\ep} = 0,
\]

\noindent as in (\ref{int}).  

Either way, Ricci--flatness is a necessary condition for the D8-brane
(\ref{soln}) to be supersymmetric.  It is not, however, a sufficient
condition since we must ensure (\ref{int}), as opposed to
(\ref{ricci}), is satisfied.  In other words, we must be able to
construct a solution of 
\be
\hat{D}_i \hat{\ep} = 0,
\label{kspinor}
\ee

\noindent a Killing spinor with respect to the metric $\ga_{ij}$.  The
trivial solution is, of course, to set $\ga_{ij} = \eta_{ij}$, in
which case $\hat{\ep}(x) = \hat{\ep}_0$ is just a constant
sixteen--component Majorana spinor.  This
flat case is the maximally supersymmetric solution, breaking one
half of the ten--dimensional supersymmetries due to the
chirality requirement (\ref{chiral}).  Other,
more general choices of $\ga_{ij}$ will, however, break a greater
fraction.  If (\ref{kspinor}) admits $N$ solutions, then $N/32$ of the
ten--dimensional supersymmetries are preserved.

It is of interest to note that all such supersymmetric solutions saturate
the usual Bogomol'nyi bound on the mass and charge densities of the brane.  To see this,
consider the supercharges per unit eight--volume
\be
Q_{\ep} = \int_{\del\S} \bar{\ep} \Ga^{\ul{a}\,\ul{b}\,\ul{c}}
\psi_{\ul{c}} d\S_{\ul{a}\,\ul{b}},
\ee

\noindent where $\S$ is a nine--dimensional space--like surface, the
integral over which
reduces to one over the one--dimensional space transverse to the brane.  The
variation of $Q_{\ep}$ is
\be
\de_{\ep_1} Q_{\ep_2} = [Q_{\ep_1}, Q_{\ep_2}] = \int_{\del\S}
N^{\ul{a}\,\ul{b}} d\S_{\ul{a}\,\ul{b}},
\label{surfaceint}
\ee

\noindent where $N^{\ul{a}\,\ul{b}} = \bar{\ep}_1
\Ga^{\ul{a}\,\ul{b}\,\ul{c}} \de_2 \psi_{\ul{c}}$ is the Nester
form~\cite{N}.  With (\ref{susya}a), we have
\be
N^{\ul{a}\,\ul{b}} = \bar{\ep}_1 \Ga^{\ul{a}\,\ul{b}\,\ul{c}} D_{\ul{c}}
\ep_2 - \quarter m e^{\frac{5}{4}\p} \bar{\ep}_1 \Ga^{\ul{a}\,\ul{b}}
\ep_2.
\ee

\noindent The surface integral (\ref{surfaceint}) is evaluated on both
sides of the domain wall, i.e. as $y \rightarrow y_0^{\pm}$.  In this
limit, we have $F(y) \rightarrow 1$, $e^{\frac{5}{4}\p} \rightarrow
1$, and $\ep \rightarrow \hat{\ep}(x)$.  Then
\be
\de_{\ep_1} Q_{\ep_2} = N^{\ul{0}\,\ul{y}} \Big|^{y=y_0^+}_{y=y_0^-},
\ee

\noindent and with $\hat{D}_i \hat{\ep}_2 = 0$, this becomes
\be
\de_{\ep_1} Q_{\ep_2} = \left. \mp \quarter m \hat{\ep}_1^{\dagger} \left(
\mathbf{1} \mp \Ga_{\ul{y}} \right) \hat{\ep}_2
\right|^{y=y_0^+}_{y=y_0^-},
\label{bogbound}
\ee

\noindent where the relative signs are set by the sign of $(y-y_0)$.
As long as we can construct Killing spinors on the worldvolume, we
thus have D8-branes with both the mass per unit volume, $M$, and the charge
per unit volume, $Z$, proportional to $m$.  Indeed, if we take
$\hat{\ep}_1 = \hat{\ep}_2 = \hat{\ep}$, then (\ref{bogbound}) becomes
$\hat{\ep}^{\dagger} \left( \mathbf{1} \mp \Ga_{\ul{y}} \right)
\hat{\ep}$ which vanishes due to the chirality requirement
(\ref{chiral}).  For an arbitrary
configuration, it can be shown that there is a Bogomol'nyi--type bound
$M \ge Z$~\cite{DGHR}; and our Ricci--flat solutions saturate this bound.  Since
all of these latter are thus BPS states, it is not the case that
the flat solution is energetically favoured over the
general Ricci--flat one.

\subsection{Supersymmetric Manifolds, Holonomy Groups and All That}

If by the D8-brane worldvolume, we mean the nine--manifold with line element
\be
d\hat{s}^2 = \ga_{ij}(x) dx^i dx^j,
\ee

\noindent then we are interested in Ricci--flat
worldvolumes which admit Killing spinors.  We therefore consider manifolds
of the form $\mathbb{R}^{n+1} \times \M^{8-n}$, with line element
\be
d\hat{s}^2 = -dt^2 + dx_1^2 + \ldots + dx_n^2 + d\td{s}^2,
\label{8branemetric}
\ee

\noindent where $d\td{s}^2$ is the line element on $\M^{8-n}$.  The spatial sections
of the D8-brane worldvolume thus have the form $\mathbb{R}^n \times
\M^{8-n}$.  We can either take this as a potential D8-brane
worldvolume or, by throwing away the $\mathbb{R}^n$ factor (and ignoring the overall
harmonic function in the bulk metric), we effectively generate
solutions which describe Ricci--flat domain walls in $D = (10-n)$:
$(8-n)$-branes, with worldvolumes of the form $\mathbb{R} \times \M^{8-n}$.
These could be dimensional reductions of the D8-brane, along the lines
of~\cite{BRGPT}, or domain walls in other types of supergravity theory.  As
we shall see below, all such domain walls can be generalised to this
Ricci--flat case, so it is indeed of relevance to study the general
D8-brane worldvolume (\ref{8branemetric}).

As is well known, the classification of
possible manifolds which admit Killing spinors is given in
terms of their holonomy groups.  Since the D8-brane admits Killing spinors as long $\M^{8-n}$
does, we are interested in manifolds $\M^d$ with
dimensions $d=8,7,6,5,4$.    The Killing spinors will then have the form
\be
\hat{\ep}(x) = \ep_0 \otimes \eta
\ee

\noindent where $\ep_0$ is a constant $SO(n,1)$ spinor and $\eta$
transforms under the holonomy group of $\M^d$.  Perusing Berger's list~\cite{Ber}, and
ignoring the trivial case, the relevant
holonomy groups are as follows.  For $d=8$, we have $H=SU(4),Sp(2)$ or
$\mbox{Spin}(7)$, the first two corresponding to K\"{a}hler and
Hyper--K\"{a}hler manifolds respectively.  Such choices will preserve
$\frac{1}{16},\frac{3}{32},\frac{1}{32}$ of the ten--dimensional
supersymmetries respectively.  For $d=7,6,4$, we have $H = G_2,SU(3),SU(2)$
respectively. (We ignore the $d=5$ case here, since the holonomy group
is then just $H=SU(2)\times \mathbf{1}$, so this effectively reduces to
the $d=4$ case.)  Manifolds with these holonomy groups will preserve
$\frac{1}{16},\frac{1}{8},\frac{1}{4}$ of the ten--dimensional
supersymmetries respectively.

In the Euclidean regime, there is a connection between supersymmetric
manifolds --- those which
admit Killing spinors --- and self--dual manifolds --- those which have
self--dual curvature two--forms~\cite{AL}.  The four--dimensional self--duality
condition on the curvature two--form can be generalised to
\be
\Theta_{\ul{a}\,\ul{b}} = \half \p_{\ul{a}\,\ul{b}\,\ul{c}\,\ul{d}}
\Theta^{\ul{c}\,\ul{d}},
\label{selfdualcurv}
\ee

\noindent where $\p_{\ul{a}\,\ul{b}\,\ul{c}\,\ul{d}}$ is a duality operator,
identified with the components of some fundamental, nowhere--vanishing
four--form on $\M^d$.  Just as in four dimensions, this second order
equation on the metric is equivalent to a first order one on the vielbein~\cite{AL}:
\be
\om_{\ul{a}\,\ul{b}} = \half \p_{\ul{a}\,\ul{b}\,\ul{c}\,\ul{d}}
\om^{\ul{c}\,\ul{d}},
\label{selfdual}
\ee

\noindent at least for some specific
choice of gauge.  Manifolds which satisfy this can
be thought of as $d$--dimensional gravitational instantons.  This is a direct
generalisation of the similar considerations for self--dual
Yang--Mills fields in $d$ dimensions: for $d > 4$, the
operator $\p$ belongs to an
irreducible representation of $SO(d)$; and if a
subgroup $H$ of $SO(d)$ can be found such that the decomposition of
this representation under $H$ contains a singlet, then the
corresponding tensor invariant $\p$
can be constructed~\cite{CDFN}.  Manifolds with self--dual
connections, in the sense of (\ref{selfdual}), will then have holonomy
group $H$.

In eight dimensions, the duality operator $\p$ can be chosen to be
invariant under one of two maximal subgroups of $SO(8)$: $(SU(4) \times
U(1))/\mathbb{Z}_4$ or $\mbox{Spin}(7)$~\cite{CDFN}.  The former
choice would generate
the eight--dimensional K\"{a}hler manifolds, although it would seem that we cannot
generate the Hyper--K\"{a}hler case using this method.  The
$\mbox{Spin}(7)$ case would
seem to be
the more interesting, however: starting with $\p$ as the unique
$\mbox{Spin}(7)$--invariant Hodge self--dual four--form, it has been shown that
any manifold which satisfies (\ref{selfdualcurv}) in
$d=8,7,6,5,4$ has holonomy group
$H=\mbox{Spin}(7),G_2,SU(3),SU(2)\times \mathbf{1},SU(2)$
respectively~\cite{AL}.  This is just the list of holonomy groups
given above.  Moreover, the $d=8$ equation is a
`master' equation from which the equations in $d < 8$ can be
derived simply by assuming $\M^8$ to be the product of $\M^{8-n}$
with $T^n$ or $\mathbb{R}^n$, since the holonomy groups of the latter
are trivial.  This exactly parallels the remarks made above: by
throwing away the $\mathbb{R}^n$ piece of the metric (\ref{8branemetric}), we
generate domain walls in lower dimensions.  The `dimensional
reduction' is the same in both cases.  So the spatial sections of the D8-brane worldvolume
can be identified with an eight--dimensional gravitational instanton,
or can be a product of flat space with a lower dimensional instanton.

Compact manifolds $\M^d$ with the required properties are well--known.  The
$d$--dimensional torus $T^d$ is the trivial example, since it has a trivial holonomy
group.  In $d=8,7$, Joyce has constructed compact manifolds with
holonomy groups $H=\mbox{Spin}(7),G_2$ respectively~\cite{J1,J2,J3}.
In six dimensions, the Calabi--Yau manifolds have
$H=SU(3)$, and in four dimensions the Hyper--K\"{a}hler manifold $K3$
has $H=SU(2)$.  All such spaces are Ricci--flat as
required.  Making use of such manifolds as the spatial sections of the
D8-brane would give a solution, the interpretation of which would be of
a D8-brane wrapped on $\M^d$.

\subsection{Explicit Non--Compact Examples}

We are more interested, however, in the \emph{un}wrapped brane, so we
turn our attention to \emph{non}--compact manifolds $\M^d$ for
$d=8,7,6,5,4$.  Starting with $d=8$, $\M^8$ can be a non--compact K\"{a}hler or
Hyper--K\"{a}hler manifold with $H=SU(4), Sp(2)$ respectively.  The
latter includes the product of two Euclidean Taub--NUT spaces, an
asymptotically locally Euclidean manifold with holonomy $H=Sp(1)
\times Sp(1)$.  

More generally, we could make use of the asymptotically locally Euclidean `toric'
Hyper--K\"{a}hler manifolds, with a tri--holomorphic $T^2$ isometry.
The eight--dimensional line element has the local form~\cite{GGPT}
\be
d\td{s}^2 = U_{AB} d\ul{x}^A d\ul{x}^B + U^{AB}
(d\p_A + A_A)(d\p_B + A_B),
\label{HK}
\ee

\noindent where $\ul{x}^A = \{x^A_s, ~ A=1,2, ~ s=1,2,3\}$
are coordinates on two copies of $\mathbb{E}^3$ and $U_{AB}$ are the entries of
a positive definite symmetric $2 \times 2$ matrix function $U$ of
these coordinates.  $U^{AB}$ are the entries of $U\inv$, and the $\p_A$ are periodically
identified with period $2\pi$.  The two
one--forms $A_A = d\ul{x}^B \cdot \ul{\om}_{BA}$, where
$\ul{\om}_{AB}$ are a triplet of $2 \times 2$ matrix functions of
the $\ul{x}^A$.  The one-forms satisfy the constraint
\be
F^{rs}_{ABC} = \vep^{rst} \del^t_A U_{BC},
\ee

\noindent where
\be
F^{rs}_{ABC} = \del^r_A \om^s_{BC} - \del^s_B \om^r_{AC},
\ee

\noindent are the components of the two--form field strength $F_A =
dA_A$.  Here
\[
\del^r_A = \frac{\del}{\del x^A_r}.
\]

A solution is~\cite{GGPT}
\be
U_{AB} = U_{AB}^{\infty} + \frac{p_A p_B}{2 \left| \sum_C p_C \ul{x}^C
- \ul{a} \right|},
\label{U}
\ee

\noindent where, if the metric is to be non--singular, $\{p_A\}$ is a
set of two coprime integers.  $\ul{a}$
is an arbitrary three--vector, specifying the `location' of a
$3$-plane in $\mathbb{E}^6$, and $U_{AB}^{\infty}$ is a constant.
The metric is entirely non--singular, as should be the case for the
D8-brane worldvolume.  More general solutions, consisting of
superpositions of (\ref{U}) can be constructed~\cite{GGPT}.  With
$U_{AB}^{\infty} = \de_{AB}$, the solution is asymptotic to $\mathbb{E}^6
\times T^2$.  With $p_A p_B = \de_{AB}$, and
$\ul{a} = \ul{0}$, the solution reduces to a product of two Euclidean
Taub--NUT manifolds, with holonomy group $H=Sp(1)\times Sp(1)$.  The
general solution, with $H=Sp(2)$, admits three $SO(8)$ Killing spinors,
so will preserve $\frac{3}{32}$ of the ten--dimensional
supersymmetries.   The isometry group in this case is just
$U(1)^2$, generated by the Killing vectors $\frac{\del}{\del \p_A}$

We turn, now, to the case in which the holonomy group of $\M^8$ is
$\mbox{Spin}(7)$.  Complete, non--compact manifolds with $H=\mbox{Spin}(7)$ have been
constructed by Gibbons \emph{et al}~\cite{GPP}; these
take the form of $\mathbb{R}^4$ bundles over $S^4$, and have the line element
\be
d\td{s}^2 = \left( 1 - \left(\frac{M}{r}\right)^{\frac{10}{3}} \right)\inv dr^2  +
\frac{9}{20}r^2 \left( d\mu^2 + \quarter \sin^2 \mu \S_s^2 \right) + \frac{9}{100} r^2
\left( 1 - \left(\frac{M}{r}\right)^{\frac{10}{3}} \right) \left(\s_s - A^s \right)^2,
\label{spin7}
\ee

\noindent where $s=1,2,3$ and $M$ is an integration constant.
The $\{\s_s\}$ and $\{\S_s\}$ are left--invariant one--forms on the
principal $SU(2)$ bundle over $S^4$, with the single--instanton
connection
\[
A^s = \cos^2 \frac{\mu}{2} \S_s.
\]

\noindent This metric is Ricci--flat as required and has the isometry
group $SO(5) \times SU(2)$.  The singularity at $r=M$ is a removable
`bolt' singularity which is topologically $S^4$.  In the limit $r
\rightarrow M$, the metric is that on $\mathbb{R}^4$, up to an overall
numerical constant, and the
boundary at infinity is the squashed seven--sphere~\cite{ADP}.  The
D8-brane with such spatial sections preserves $\frac{1}{32}$ of the
ten--dimensional supersymmetries, since the $\mbox{Spin}(7)$ manifold
admits a single Killing spinor.

The fact that (\ref{spin7}) is a self--dual manifold has been underlined
in~\cite{BFK}: the same metric is a solution of the
self--duality condition (\ref{selfdual}), with
$\p_{\ul{a}\,\ul{b}\,\ul{c}\,\ul{d}}$ the components of the unique Hodge
self--dual $\mbox{Spin}(7)$--invariant four--form.  In this case, the
self--duality conditions become~\cite{CDFN,BFK}
\be
\om_{8\ul{a}} = \half c_{\ul{a}\,\ul{b}\,\ul{c}} \om^{\ul{b}\,\ul{c}},
\label{8dselfdual}
\ee

\noindent where now $a,b = 1, \ldots 7$ and $c_{\ul{a}\,\ul{b}\,\ul{c}}$
are the octonionic structure constants.  The D8-brane with
such a worldvolume thus has spatial
sections which are eight--dimensional gravitational instantons.  Since $S^7$ can
be thought of as an $S^3$ bundle over $S^4$, and $S^3 = SU(2)$, we see
why the $SU(2)$ connection appears in the metric.  In some sense, this
manifold is the eight--dimensional generalisation of
the Eguchi--Hanson space, the latter making use of the fact that $S^3$ is an
$U(1)$ bundle over $S^2$, and having a squashed three--sphere as its
boundary at infinity.

\section{7-Brane Domain Walls with Holonomy $G_2$}

As explained in~\cite{BRGPT}, the Romans theory
can be dimensionally reduced to generate a massive nine--dimensional
supergravity theory, despite the fact that the product of
nine--dimensional Minkowski space with a circle is not a solution of
the Romans theory.  All one needs is a solution with a $U(1)$
isometry, and this is provided by the D8-brane solution.  The new
nine--dimensional theory has a `cosmological constant' in exactly the
same way as the Romans theory does; and this allows for the existence of
a 7-brane domain wall solution --- the double dimensional reduction of the
D8-brane.  Moreover, it was shown
in~\cite{BRGPT} that this 7-brane solution is T--dual to the direct
dimensional reduction of the D7-brane of type IIB supergravity,
although the dimensional reduction in this case must be of a
Scherk--Schwarz type~\cite{SS}.

To discuss the 7-brane domain wall of this massive nine--dimensional
supergravity theory consider, first, a D8-brane solution of the form
\be
ds^2 = F^2 \left( \ga_{ij}(x) dx^i dx^j + dy_1^2 \right) + dy_2^2,
\ee

\noindent where now $\ga_{ij}$ is the metric on an eight--dimensional
manifold.  By taking
$y_1$ as the coordinate on a circle, and with $F(y_2)$ and the dilaton
as in (\ref{soln}), this is the generalisation of the dimensionally
reduced D8-brane of~\cite{BRGPT}.  By throwing away
the circular dependance, we have 7-branes with general
worldvolumes.  These are solutions of the massive nine--dimensional
theory of~\cite{BRGPT} if and only if the metric $\ga_{ij}$ is
Ricci--flat, exactly as in the case above.  If the 7-brane worldvolume of the form
\be
d\hat{s}^2 =  \ga_{ij}(x) dx^i dx^j = -dt^2 + d\td{s}^2,
\ee

\noindent is to be supersymmetric, the holonomy group of $\M^7$, with
line element $d\td{s}^2$, must be
$H=G_2$.  Such manifolds have again been constructed by Gibbons
\emph{et al}~\cite{GPP}, and
have a form similar to the $\mbox{Spin}(7)$ manifold (\ref{spin7}) discussed above.
That is, $\M^7$ can be an $\mathbb{R}^3$ bundle over $S^4$, or an
$\mathbb{R}^4$ bundle over $S^3$.

With an appropriate duality operator, and with the caveat to be
discussed below, these manifolds solve the
seven--dimensional self--duality condition, just as the $\mbox{Spin}(7)$ manifold (\ref{spin7}) does in eight dimensions.  To see
this, consider the self--duality condition (\ref{selfdual}) with
$\p_{\ul{a}\,\ul{b}\,\ul{c}\,\ul{d}}$ the
components of the $G_2$--invariant, seven--dimensional Hodge dual of the
octonionic structure constants.  Explicitly~\cite{FK},
\be
\p_{\ul{a}\,\ul{b}\,\ul{c}\,\ul{d}} = \frac{1}{3!}
\vep_{\ul{a}\,\ul{b}\,\ul{c}\,\ul{d}\,\ul{e}\,\ul{f}\,\ul{g}}
c^{\ul{e}\,\ul{f}\,\ul{g}},
\ee

\noindent where
\be
c_{\ul{a}\,\ul{b}\,\ul{c}} = +1 ~~~~ {\rm for} ~~~~
\ul{a}\,\ul{b}\,\ul{c} = 123, 516, 624, 435, 471, 673, 572.
\ee

\noindent We thus have
\be
\p_{\ul{a}\,\ul{b}\,\ul{c}\,\ul{d}} = +1 ~~~~ {\rm for} ~~~~
\ul{a}\,\ul{b}\,\ul{c}\,\ul{d} = 1245, 2671, 3526, 4273, 5764, 6431,
7531.
\ee

\noindent The self--duality conditions (\ref{selfdual}) are then
\begin{eqnarray}
\om_{71} &=& \om_{26} + \om_{53}, ~~~~~~~  \om_{72} = \om_{61} +
\om_{34}, ~~~~~~~ \om_{73} = \om_{42} + \om_{15},\nonumber\\
\om_{74} &=& \om_{23} + \om_{65}, ~~~~~~~ \om_{75} = \om_{46} + \om_{31},
~~~~~~~  \om_{76} = \om_{12} + \om_{54},\nonumber\\
\om_{63} &=& \om_{25} + \om_{14}.
\label{g2spinconn}
\end{eqnarray}

With the ansatz
\be
d\td{s}^2 = f^2(r) dr^2 + g^2(r) \S_s^2 + h^2(r) \left( \s_s - A^s \right)^2,
\label{g2}
\ee

\noindent the metric on $\M^7$ has the form of an $\mathbb{R}^4$
principal bundle over $S^3$.  Since $S^3$ is parallelizable, this
bundle is trivial.  $\{\S_s\}$ are the left--invariant one--forms on the
base space and $\{\s_s\}$ the left--invariant one--forms on the fibres.  They satisfy
\[
\S_s = -\half \vep_{stu} \S_t \wedge \S_u, ~~~~~~~ \s_s =
-\half \vep_{stu} \s_t \wedge \s_u.
\]

\noindent $A^s$ is the connection on the bundle, given by
\be
A^s = \half \S_s.
\ee

\noindent With the orthonormal one--forms $e^{7} = f dr$,
$e^{\ul{s}} = g \S_s$, and $e^{\ul{\hat{s}}} = h \left( \s_s - A^s
\right)$, where $\ul{\hat{s}} = 4,5,6 = \hat{1},\hat{2},\hat{3}$, the
connection one--forms are given by
\[
\om^{\ul{s}}_{~\,7} = \frac{g'}{fg} e^{\ul{s}}, ~~~~~~~
\om^{\ul{\hat{s}}}_{~\,7} = \frac{h'}{fh} e^{\ul{\hat{s}}},~~~~~~~
\om^{\ul{s}}_{~\,\ul{\hat{t}}} = \frac{1}{8}\frac{h}{g^2} \vep_{stu}
e^{\ul{u}},
\]
\be
\om^{\ul{s}}_{~\,\ul{t}} = -\half \frac{1}{g} \vep_{stu} \left(
e^{\ul{u}} + \quarter \frac{h}{g} e^{\ul{\hat{u}}} \right), ~~~~~~
\om^{\ul{\hat{s}}}_{~\,\ul{\hat{t}}} = -\half \frac{1}{h} \vep_{stu}
\left( e^{\ul{\hat{u}}} + \frac{h}{g} e^{\ul{u}} \right).
\ee

\noindent Substituting into the self--duality conditions
(\ref{g2spinconn}), we find the following first order differential
equations
\begin{subeqnarray}
&& \quarter \frac{h}{g^2} + \frac{g'}{fg} = 0,\label{dea}\\
&& \half \frac{1}{h} - \frac{1}{8} \frac{h}{g^2} + \frac{h'}{fh} = 0, \label{deb}
\end{subeqnarray}

\noindent Using the reparametrisation invariance of the metric
(\ref{g2}) under
$r \rightarrow r' = r'(r)$, we can set
\be
g^2(r) = \frac{1}{12} r^2,
\ee

\noindent in which case, the solution of (\ref{dea}a,b) is
\be
f^2(r) = \left( 1 - \left( \frac{M}{r} \right)^3 \right)\inv, ~~~~~~ h^2(r) =
\frac{1}{9} r^2 \left( 1 - \left( \frac{M}{r} \right)^3 \right),
\ee

\noindent where $M$ is an arbitrary integration constant.  This is
precisely the metric found by solving the Einstein equations
in~\cite{GPP}; here, we have derived it from first order equations
alone.  In direct analogy with the $\mbox{Spin}(7)$ case (\ref{spin7}) above,
the `bolt' singularity here is topologically $S^3$, and as $r \rightarrow
M$, the metric (\ref{g2}) reduces to the metric on $\mathbb{R}^4$.
The boundary at infinity is $S^3 \times S^3$ and the isometry
group is $SO(4) \times SU(2)$.

The other seven--dimensional $G_2$ metric considered by Gibbons
\emph{et al}~\cite{GPP} is that of an $\mathbb{R}^3$ bundle over $S^4$.  This is
somewhat more complicated than the above example since the bundle is
no longer trivial.  It can be shown that $G_2$ holonomy implies
self--dual curvature in the above ($G_2$) sense and, for this reason,
it should be expected that the $\mathbb{R}^3$ bundle over $S^4$ is
self--dual in the same way as the $\mathbb{R}^4$ bundle over $S^3$.
This is not the case, however, at least na\"{\i}vely.  That is,
although the $\mathbb{R}^3$ bundle over $S^4$ is Ricci--flat and has $G_2$ holonomy,
its connection one--form does not satisfy the relations
(\ref{g2spinconn}).  Moreover, neither does its curvature two--form
satisfy the same relations with $\om_{\ul{a}\,\ul{b}}$ replaced by
$\Theta_{\ul{a}\,\ul{b}}$.  A possible explanation of this is as follows.
Transforming the orthonormal basis by an arbitrary $SO(7)$ rotation
will leave the line element unchanged.  Such a rotation will have an
effect on the curvature two form however: schematically, if $e \rightarrow G
e$, where $G \in SO(7)$, we have $\Theta \rightarrow G^T \Theta G$.
Then the question is whether this transformation commutes with the
duality operator $\half \p$ in (\ref{selfdualcurv}).  Indeed, since the
duality operator is $G_2$--invariant, and not $SO(7)$--invariant,
it is unlikely to commute with such a rotation.  It would seem, then,
that a judiciously chosen $SO(7)$ rotation of the
basis will ensure self--duality of the curvature two--form.  This is
similar to the four--dimensional case in which, via an $SO(4)$
rotation, a manifold with self--dual curvature can always be brought
into a form such that its connection is also self--dual; the
difference being, of course, that here we do not have a self--dual
curvature in the first place.

At any rate, since both the $\mathbb{R}^4$ bundle over $S^3$ and the $\mathbb{R}^3$
bundle over $S^4$ are Ricci--flat manifolds of holonomy $H=G_2$,
either manifold can be used as the spatial sections of a
supersymmetric 7-brane
in nine dimensions.

\section{The General Domain Wall}

Considerations similar to the above hold for domain walls in any
supergravity theory.  For example, we can dimensionally reduce the
D8-brane along the lines of~\cite{BRGPT}, generating domain walls in
$D < 10$ massive supergravity theories.  Or we can consider domain
walls in massive gauged supergravity theories~\cite{LPSS}.

In $D$ dimensions, the domain wall is a
$(D-2)$-brane, which couples to a potential, or `cosmological constant'
\be
V(\p) = 2\Lambda e^{\al \p}.
\ee

\noindent $\al$ is a constant which can be parametrised as
\be
\al^2 = \De + 2 \frac{(D-1)}{(D-2)}.
\label{alpha}
\ee

\noindent The value of $\De$ varies from case to case~\cite{LPT}.  In certain
vacua of gauged supergravities, we
have $\De = -2(D-1)/(D-2)$, in which case $\al = 0$, and the potential
is a true cosmological constant.  On the other hand, some gauged seven-- and four--dimensional
supergravities have $\De = -2$.  We will take $\De = 4$, however,
since this is the value of $\De$ for the Romans theory and all its
dimensional reductions.  (Incidentally, the Howe--Lambert--West
massive supergravity theory~\cite{HLW} has $\De=0$.  It would
seem, then, that this theory does not have the usual brane solutions, as in
(\ref{intro}).)

The relevant Lagrangian is
\be
\L = \sqrt{-g} \left( R - \half |\del \p|^2 - V(\p) \right),
\ee

\noindent the equations of motion of which are
\begin{subeqnarray}
R_{ab} &=& \half \del_a \p \del_b \p + \frac{V(\p)}{D-2}g_{ab}, \label{eomdwalla}\\
D^2 \p &=& \frac{dV}{d\p}.\label{eomdwallb}
\end{subeqnarray}

\noindent These have the domain wall solution~\cite{LPT}
\be
\left. \begin{array}{ccl} ds^2 &=& H^{\frac{1}{D-2}} d \ul{x} \cdot d \ul{x} +
H^{\frac{D-1}{D-2}}dy^2,\\ [.07in]
e^{\p} &=& H^{-\frac{\al}{2}}, \end{array} \right\}
\ee

\noindent where $H(y)$ is harmonic on the single transverse
direction.  As explained in~\cite{LPT}, we can take 
\be
H(y) = 1 + m |y - y_0|,
\ee

\noindent to avoid potential curvature singularities, as for the
D8-brane above.

Keeping the dilaton field and the specific form of the harmonic
function $H(y)$ as above, we generalise the metric to
\be
ds^2 = H^{\frac{1}{D-2}} \ga_{ij}(x) dx^i dx^j +
H^{\frac{D-1}{D-2}}dy^2.
\ee

\noindent This is a solution of the equations of motion
(\ref{eomdwalla}a,b), as long as the metric $\ga_{ij}$ is Ricci--flat.
That is, any domain wall --- not just the D8-brane of
above --- can in general have a Ricci--flat worldvolume.

The remarks concerning supersymmetry that we made above can be applied
here.  That is, the general domain wall will preserve supersymmetry if
and only if the
worldvolume manifold with line element
\[
d\hat{s}^2 = \ga_{ij}(x) dx^i dx^j,
\]

\noindent admits Killing spinors.  As already mentioned, the
dimensional reduction of the Romans
theory leads to a massive supergravity theory in nine dimensions, with
a 7-brane domain wall.  The worldvolume metric of this can then have spatial
sections with holonomy group $H=G_2$.  In $D=8$, we can have a 6-brane
supersymmetric domain wall, the
spatial sections of which take the form of non--compact Calabi--Yau
manifolds with holonomy group $H=SU(3)$~\cite{C}.

Moving down in dimension, we can have 4-brane domain walls in $D=6$
supergravity theories.  These solutions will be supersymmetric if the
spatial sections of their worldvolumes have holonomy group $H=SU(2)$.
In the compact case, we could consider
\[
d\hat{s}^2 = -dt^2 + d\td{s}^2_{K3},
\]

\noindent where $d\td{s}^2_{K3}$ is the metric on $K3$.  This has the interpretation of a
4-brane wrapped on $K3$.  Or, in the more interesting non--compact case,
\[
d\hat{s}^2 = -dt^2 + d\td{s}^2_{TN},
\]

\noindent where $d\td{s}^2_{TN}$ is the asymptotically locally Euclidean Taub--NUT
metric.  The 4-brane worldvolume is then just the five--dimensional
Kaluza--Klein monopole~\cite{GP,S}.

An interesting example is that of the domain walls in
Ho\v{r}ava--Witten theory.  As is well known, the strongly coupled
$E_8 \times E_8$ heterotic string theory is just M-Theory compactified
on an $S^1/\mathbb{Z}_2$ orbifold with a set of $E_8$ gauge fields on each of
the orbifold fixed planes~\cite{HW1,HW2}.  The compactification of the
eleven--dimensional theory on a Calabi--Yau manifold leads to a gauged
five--dimensional supergravity theory with two four--dimensional
boundaries.  This has a solution describing
two parallel supersymmetric 3-brane domain walls located at the
orbifold fixed planes~\cite{LOSW}.  On a further dimensional
reduction, four--dimensional spacetime is identified with the 3-brane
worldvolume --- the universe as a domain wall scenario~\cite{LOSW}.
This being a specific example of the general case considered here, it
should be obvious that the usual flat 3-brane domain walls can be
generalised to any Ricci--flat worldvolume.

The relevant five--dimensional Lagrangian is~\cite{LOSW}
\begin{eqnarray}
S &=& S_{\rm bulk} + S_{\rm boundary},\nonumber\\
S_{\rm bulk} &=& \half \int_{\M^5} \sqrt{-g} \left( R -
\half \frac{1}{V^2} | \del V |^2 - \third \frac{1}{V^2} \al^2
\right),\\
S_{\rm boundary} &=& \sqrt{2} \left(
\int_{\M^4_{(1)}} \sqrt{-g} \frac{\al}{V} - \int_{\M^4_{(2)}}
\sqrt{-g} \frac{\al}{V} \right),\nonumber
\end{eqnarray}

\noindent where $V$ is a modulus which encodes the variation of the
Calabi--Yau volume and $\al$ is a constant `mass term', for the definition of
which we refer the reader to~\cite{LOSW}.  The equations of motion will admit 3-brane
solutions of the form
\be
\left.\begin{array}{ccl} ds^2 &=& H \ga_{ij}(x) dx^i dx^j + H^4 dy^2,\\ [.07in]
V(y) &=& H^3,\\ [.07in]
H(y) &=& 1 + \frac{\sqrt{2}}{3} \al |y|,
\end{array}\right\}
\ee

\noindent if and only if $\ga_{ij}$ is Ricci--flat.  (In the above we
have set three arbitrary constants to unity.)  We have checked this
solution explicitly.

These 3-branes can have worldvolumes that are
four--dimensional Schwarzschild black holes.  Although not
supersymmetric, since the Schwarzschild solution does not admit
Killing spinors, this is of interest nonetheless.  It explains how our
four--dimensional universe, upon a further dimensional reduction,
could have the form of a Schwarzschild black hole.  It would seem
that this embedding of the Schwarzschild solution in five dimensions violates the no--go
theorem of~\cite{KSS} but in fact it does not.  The theorem is that it is
impossible to embed the Schwarzschild solution in a \emph{flat}
five--dimensional spacetime; and our five--dimensional manifold is
certainly not flat, so there is no contradiction.

\section{Magnetic Branes in General}

There has been some discussion in the literature of solutions of
supergravity theories which describe generalised electric branes, in
which the transverse space is no longer flat.  Such solutions preserve
fewer than the usual one half of the spacetime supersymmetries.  The
eleven--dimensional membrane solution can be generalised so as to
interpolate between eleven--dimensional Minkowski space and $AdS_4
\times \M^7$, where $\M^7$ is any Einstein space~\cite{DLPS}; the squashed
seven--sphere for example, which admits Killing spinors
and so allows for generalised supermembranes.  Indeed, the
eight--dimensional $\mbox{Spin}(7)$ and Hyper--K\"{a}hler manifolds
(\ref{spin7}) and (\ref{HK}) above have also been considered as
possible transverse spaces of the eleven--dimensional
supermembrane~\cite{BFK,GGPT}; or as the transverse space of the
fundamental string in ten dimensions~\cite{BFK}.  Here we show that the
worldvolume, as opposed to the transverse space, of the magnetic
branes can also be generalised in just such a manner.

A general magnetic $p$-brane, with $p=D-n-2$ couples to a rank $n$
field strength $F_{[n]}$.  The domain wall is
then a specific example of the more general magnetic brane, one with
$n=0$ (or, by Hodge duality, $n=D$)~\cite{LPT}.  This leads
us to consider whether we can generalise the worldvolume metric of any
magnetic brane in the above manner; and the answer is that we can.
Heuristically, since the field strength which couples to a magnetic brane
is non--zero in the transverse directions only and, since we change
the worldvolume metric only, the generalisation will go through in
precisely the same way as above.  Indeed, it was noted in~\cite{K}
that the M5-brane worldvolume can be generalised to
\be
d\hat{s}^2 = -dt^2 + dx^2 + ds^2_{K3}.
\label{5branek3}
\ee

\noindent This solution has the
interpretation of an M5-brane wrapped on $K3$, as considered from the
point of view of the worldvolume action in~\cite{SC}.  It is a
specific example of the more general claim that we are making here.

The Lagrangian relevant to the study of a general magnetic brane is
\be
\L = \sqrt{-g} \left( R - \half \del \p \cdot \del \p - \frac{1}{2(n!)}
e^{\al \p} F^2_{[n]} \right),
\label{lmagbrane}
\ee

\noindent where the dimensionality of the brane is given by $d = p+1 =
D-n-1$, and where $\al$ is as in (\ref{alpha}) (we set $\De = 4$
here).  $F_{[n]}$ is the field strength which couples to the
$p$-brane.  The equations of motion
\begin{subeqnarray}
&& R_{ab} = \half \del_a \p \del_b \p + \frac{1}{2(n-1)!} e^{\al \p} \left(
F_{a a_1 \ldots a_n} F_b^{~a_1 \ldots a_n} - \frac{(n-1)}{n(D-2)} F^2 g_{ab}
\right),\label{pbraneeoma}\\
&& D^2 \p = \frac{\al}{2n!}e^{\al \p}F^2,\label{pbraneeomb}\\
&& D_{a} \left (e^{\al \p} F^{a a_1 \ldots a_n} \right) = 0,\label{pbraneeomc}
\end{subeqnarray}

\noindent have the solution
\be
\left. \begin{array}{ccl} ds^2 &=& H^{-\frac{\td{d}}{(D-2)}} \ga_{ij}(x) dx^i dx^j +
H^{\frac{d}{(D-2)}} d\ul{y} \cdot d\ul{y},\\ [.07in]
e^{\p} &=& H^{-\frac{\al}{2}},\\ [.07in]
F_{\al_1 \ldots \al_n} &=& \la \vep_{\al_1 \ldots \al_n
\beta}\frac{y^\beta}{r^{n+1}},\end{array}\right\}
\label{magbrane}
\ee

\noindent as long as the worldvolume metric $\ga_{ij}$ is
Ricci--flat.  Here, $\{y^{\al}\}$ are the coordinates on the
transverse space, $r$ being the radial coordinate in these
directions and $H(r)$ is, as usual, harmonic on this space.  $\td{d} = D-d-2=n-1$ is the
dimensionality of the electric brane dual
to the magnetic one and the alternating
tensor in the expression for the field strength has components $\pm1$.

Consider the prototypical example: the M5-brane.  The general solution
of the eleven--dimensional equations of motion is
\be
\left. \begin{array}{ccl} ds^2 &=& H^{-\third} \ga_{ij}(x) dx^i dx^j + H^{\frac{2}{3}} d\ul{y}
\cdot d\ul{y},\\ [.07in]
F_{\al\beta\ga\de} &=& \pm 3k \vep_{\al\beta\ga\de\ep}
\frac{y^{\ep}}{r^5},\\ [.07in]
H(r) &=& 1 + \frac{k}{r^3},\end{array} \right\}
\label{5brane}
\ee

\noindent where $k$ can be related to the tension of the brane by the
inclusion of a source term in the action.  These solutions are
supersymmetric if and only if the worldvolume admits Killing spinors,
just as for the cases considered above.  The supersymmetry transformation
of the gravitino is
\be
\de \psi_a = \left( D_a  - \frac{1}{288} \left( \Ga_a^{~\,bcde} - 8
\de_a^b \Ga^{cde} \right) F_{bcde} \right) \ep,
\ee

\noindent where $\ep$ is an arbitrary $32$--component Majorana
spinor.  We make the usual $6+5$ split
\be
\Ga_{\ul{a}} = \left( \ga_{\ul{i}} \otimes \S , \mathbf{1} \otimes
\S_{\ul{\al}} \right),
\ee

\noindent where $\ga_{\ul{i}}$ and $\mathbf{1}$ are $SO(5,1)$
matrices, $\S$ and $\S_{\ul{\al}}$ are $SO(5)$ matrices, $\S =
\S_6 \ldots \S_{10}$, so that $\S^2 = \mathbf{1}$, and we take
\be
\ep(x,y) = \hat{\ep}(x) \otimes \eta(r).
\ee

\noindent Substituting for
the solution (\ref{5brane}), and setting the
variation $\de \psi_{\al} = 0$ gives
\be
\eta(r) = H^{-\frac{1}{12}} (r) \eta_0,
\ee

\noindent for a constant spinor $\eta_0$, in addition to the usual chirality condition
\be
\left( \mathbf{1} \mp \S \right) \eta = 0,
\ee

\noindent which removes one half of the supersymmetries.  The
remaining condition, $\de\psi_i = 0$, is satisfied if and only if
\be
\hat{D}_i \hat{\ep} = 0,
\ee

\noindent as promised.  All solutions which satisfy this condition
will saturate the usual Bogomol'nyi bound, as in the D8-brane case above.

Instead of the compact manifold in (\ref{5branek3}), we can make use
of the asymptotically locally Euclidean Taub--NUT metric, giving a worldvolume of
the form
\be
d\hat{s}^2 = -dt^2 + dx^2 + ds^2_{TN},
\label{5branetn}
\ee

\noindent which is supersymmetric since the holonomy group of the
Taub--NUT space is $H=SU(2)$.  There do not seem to be many other
possibilities.  The Ricci--flat M5-branes no longer
interpolate between eleven--dimensional Minkowski space and $AdS_7
\times S^4$.  For a worldvolume of the form (\ref{5branetn}), the
solution at infinity in the transverse space is ${\rm Mink}_7 \times$
Taub--NUT, i.e. the Kaluza--Klein monopole oxidised to eleven
dimensions.  As $r \rightarrow 0$, the line element is
\be
ds^2 = R^2 ( -dt^2 + dx^2 ) + 4k^{\frac{2}{3}} \frac{dR^2}{R^2} + R^2
ds^2_{TN} + k^{\frac{2}{3}} d\Om^2_4,
\ee

\noindent which is the metric on the warped product of Taub--NUT with
$AdS_3 \times S^4$, where the $AdS_3$ piece has cosmological
constant $\Lambda = -1/(4k^{2/3})$.

The above considerations can be applied to any magnetic brane,
although the specific form of the supersymmetric solutions must be
considered case by case.

\subsection{Worldvolume Solitons}

Asymptotically, the spacetime metrics of the usual
brane solutions (\ref{intro}) are flat and it is well--known that both
\Ds~and M-branes admit worldvolume
solitons when embedded in such a flat
spacetime~\cite{GGT}.  That is, the Dirac--Born--Infeld (DBI) worldvolume
lagrangian is linearised and the energy of the branes is minimised for
solitonic configurations of the worldvolume fields.  Now the more
general solutions we have been discussing have a Ricci--flat spacetime
metric at infinity; and it is of interest
to note that branes embedded in such Ricci--flat spacetimes also have
worldvolume solitons, as long as
an (anti--)self--dual gauge field can be constructed on the
worldvolume manifold.

Consider, then, the simplest example, that of the D4-brane, the double
dimensional reduction of the M5-brane considered above.  At infinity,
the spacetime line element has the form
\be
ds^2 = -dt^2 + \ga_{AB} dx^A dx^B + d\ul{y} \cdot d\ul{y} = -dt^2 +
ds^2_{TN} + d\ul{y} \cdot d\ul{y},
\label{spacetime}
\ee

\noindent where $A,B = 1, \ldots 4$ and $ds^2_{TN}$ is the
asymptotically locally Euclidean Taub--NUT
metric, given by
\be
\left. \begin{array}{ccl} ds^2_{TN} &=& M^2 V\inv ( d\psi \pm
\cos\theta d\p )^2 + V (dR^2 + R^2 d\Om^2),\\ [.07in]
V(R) &=& 1 + \frac{M}{R}. \end{array} \right\}
\label{TN}
\ee

\noindent The plus (minus) sign corresponds to a self--dual
(anti--self--dual) manifold respectively and $M$ is an integration constant.

We work in an orthonormal basis, take all
worldvolume scalars to be constant and split the worldvolume coordinates
$\xi^i=\{\xi^0,\xi^A\}$.  For static configurations, $F_{0A} =
0$, and $-\det(g_{ij} + F_{ij}) = (-g_{00}) \det(g_{AB} + F_{AB})$,
where $g_{ij}$ is the pullback of the spacetime metric.  The DBI
lagrangian has the form~\cite{GGT}
\begin{eqnarray}
\L &=& e \left( 1 - \sqrt{ 1 + \quarter F^2 + \quarter \td{F}^2 +
\frac{1}{16} (F \cdot \td{F})^2} \right) \nonumber\\
&=&  e \left( 1 - \sqrt{ \left(1 \pm \quarter F \cdot \td{F}
\right)^2 - \quarter \mbox{tr} \left| F \mp \td{F} \right|^2 } \right),
\end{eqnarray}

\noindent where we have set the brane tension and the inverse string tension
$2\pi\alpha'$ to unity.  $e$ is the determinant of $e^{\ul{A}}_{~A}$, the
vierbein of $g_{AB}$.  $F^2
= F_{\ul{A}\,\ul{B}} F^{\ul{A}\,\ul{B}}$, $\td{F}$ is the Hodge dual of $F$ with respect to
the worldspace directions and $F \cdot \td{F} =
F_{\ul{A}\,\ul{B}} \td{F}^{\ul{A}\,\ul{B}}$.  The
lagrangian is linearised for configurations which satisfy
$F_{\ul{A}\,\ul{B}}=\pm\td{F}_{\ul{A}\,\ul{B}}$
and, since we are dealing with the purely static case the energy
density, $T^{00}=-\L$, is minimised for such (anti--)self--dual field
strengths~\cite{GGT}.  D4-branes embedded in a general Ricci--flat
spacetime will thus have worldvolume solitons --- in this case,
abelian (anti--)instantons --- if and only if a field strength which
is (anti--)self--dual with respect to $g_{AB}$ can be constructed.

Since no worldvolume
scalars are excited, the spatial component of the pullback of the spacetime
metric in the static gauge is just $g_{AB} = \ga_{AB}$.  The gauge field which gives an
(anti--)self--dual field strength with respect to this metric is
well--known.  With the orthonormal one--forms $e^0 = V^{1/2} dR$, $e^1 =
V^{1/2} R d\theta$ and $e^2 = V^{1/2} R \sin \theta d\p$, $e^3 = M V^{-1/2} ( d\psi \pm
\cos\theta d\p )$, it is~\cite{EH}
\be
A = \frac{1}{M} V^{-1/2} e^3,
\ee

\noindent which has the field strength
\be
F = \frac{1}{(R+M)^2}(e^0 \wedge e^3 \mp e^1 \wedge e^2).
\ee

\noindent This is manifestly (anti--)self--dual.  Although the gauge
field has the usual string--like singularity along
the $z$--axis, the energy of the instanton is finite.

D4-branes embedded in the spacetime (\ref{spacetime}) thus have
worldvolume solitons in the same way that the standard flat
D4-branes do, since (anti--)self--dual gauge fields can still be
constructed on their spatial sections.  This is of interest for two reasons.
Firstly, it should be the case that the M5-brane with worldvolume
metric (\ref{5branetn}) discussed
above will have worldvolume string solitons along similar lines
as for the flat case~\cite{HLW2}.  Secondly, it should be possible to generalise our
reasoning to the non--abelian DBI action~\cite{T,BP} describing
multiple D4-branes.  In particular, since it is known how to construct
(anti--)self--dual $SU(2)$ gauge fields on Taub--NUT space~\cite{CD}, we could
consider the action describing two D4-branes with a two--centered Taub--NUT space as
the spatial sections of the worldvolume.  The energy of such configurations will be
minimised by these more general non--abelian instantons is precisely the same way as for
the flat case~\cite{B}.  This might provide some clues as to the
nature of \Ds~in curved spacetime, a subject which is far from being
fully understood.

\section{Conclusions}

We have shown that the worldvolumes of all the magnetic branes of
string theory can take the
form of any Ricci--flat manifold.  A specific case is the domain wall,
the ten--dimensional example
being that of the D8-brane of string theory.  We have shown in detail
that the D8-brane solution of Romans' massive
type IIA supergravity theory can be generalised to include a large
class of possible solutions.  These are supersymmetric if and only if the
worldvolume manifold admits Killing spinors, a necessary, but not
sufficient, requirement for which is Ricci--flatness.

We have described some of the eight--dimensional manifolds in the
literature which do admit Killing spinors, and elaborated on the
fact that these are just the self--dual eight--dimensional
gravitational instantons.  The dimensional reduction
of the D8-brane is a seven--dimensional domain wall in nine
dimensions.  In this case, we have shown that the
known manifolds which admit Killing spinors satisfy the
self--duality condition in seven dimensions, although the connection
between supersymmetric and self--dual manifolds would seem to be a
subtle one in certain cases.

The fact that the magnetic branes can have Ricci--flat worldvolumes is
perhaps to be expected,
given the similar results concerning the transverse spaces of the
electric branes: roughly speaking, the transverse space of an electric
brane is interchangeable with the worldvolume of the dual
magnetic brane.  More speculatively, perhaps the requirement of
Ricci--flatness in these cases is a consequence of the beta functions
of string theory.  After all, to first order in the inverse string tension
$\al'$, the beta functions of the conformal field theory on the closed
string worldsheet imply that the ambient spacetime must be
Ricci--flat~\cite{CFMP}.\newline\newline

\noindent {\bf Acknowledgments}

One of the authors (DB) would like to thank Andrew Chamblin, Steve
Hewson, Jan Gutowski and Harvey Reall for various discussions.

\end{document}